\documentclass[10pt]{article}
\usepackage[left=2.5cm,top=2.5cm,right=2.5cm,bottom=1.5cm]{geometry}
\usepackage{graphicx}
\usepackage{latexsym}
\usepackage{epstopdf}
\usepackage{amsmath}
\DeclareGraphicsExtensions{.eps}
\begin{document}
\begin{center}
\large{\bf{Bianchi type-III THDE quintessence model with hybrid expansion law}} \\
\vspace{10mm}
\normalsize{ Gunjan Varshney$^1$, Anirudh Pradhan$^2$, Umesh Kumar Sharma$^3$ }\\
\vspace{5mm}
\normalsize{$^{1,2,3}$Department of Mathematics, Institute of Applied Sciences \& Humanities, GLA University,\\
Mathura-281 406, Uttar Pradesh, India \\
\vspace{2mm}
$^1$ E-mail:gunjanvarshney03@gmail.com\\
\vspace{2mm}
$^2$E-mail:pradhan.anirudh@gmail.com\\
\vspace{2mm}
$^3$E-mail:sharma.umesh@gla.ac.in} \\

\end{center}
\vspace{10mm}
\begin{abstract}
The current research investigates the behavior of the Tsallis holographic dark energy (THDE) model with quintessence in a homogeneous and 
anisotropic Bianchi type-III (B-III) space-time. We construct the model by using two conditions (i) expansion scalar ($\theta$) is proportionate
to shear scalar ($\sigma$) in the model and (ii) hybrid expansion law $a = t^\beta e^{\gamma t}$, where $\beta>0$, $\gamma>0$. Our study is 
based on Type Ia supernovae (SNIa) data in combination with CMB and BAO observations (Giostri et al, JCAP 3, 27 (2012), 
arXiv:1203.3213v2[astro-ph.CO]), the present values of Hubble constant and deceleration parameter are $H_{0} = 73.8$ and 
$q_{0} = -0.54$ respectively. Compiling our theoretical models with this data, we obtain $\beta = 2.1445~ \& ~ 2.1154$ for 
$\gamma = 0.5 ~ \& ~ 1$ respectively. We have completed a new type of cosmic model for which the expansion occurs to the current 
accelerated phase for the restraints. We have discussed the conformity among the scalar field model of quintessence and THDE model. To understand 
the Universe, we have also established the relations for Distance modulus, Luminosity Distance, and Angular-diameter distance. Some geometric 
and physical aspects of the THDE model are also highlighted.

 \end{abstract}

\section{Introduction}
Before 1998, it was ordinarily anticipated that the Universe is either in the expanded phase with a constant rate or the expansion of 
the Universe was diminishing. In 1998, the unexpected revelation of the Universe that there is an accelerated expansion in the Universe 
based on SNIa (type Ia supernova) pushed the researchers to revise the numerous cosmological model introduced so far. At present time 
existing accelerated enlargement of the cosmos is a debatable issue for physicists. Through the last three decades, the newest verdicts 
on the observational area by numerous cosmic missions like observations on large scale structure (LSS) analysis \cite{ref1,ref2}, 
Chandra X-ray observatory \cite{ref3}, BOSS collaboration\cite{ref4}, type Ia Supernovae (SNe Ia) \cite{ref5}$-$\cite{ref9}, WMAP 
collaboration \cite{ref10}, SDSS collaboration \cite{ref11,ref12}, the Hubble space telescope cluster supernova survey $V$ \cite{ref13}, 
CMBR fluctuations \cite{ref14,ref15}, latest Planck collaboration results \cite{ref16}, and the WiggleZ dark energy survey \cite{ref17}, 
confirms that our Universe is undergoing by an accelerated expansion mode. Before emerging in the late-time and current, accelerated era, 
the universe experienced an early-time accelerated phase, accompanied by the matter-dominated and the sequence of radiation represented 
by the concordance paradigm of cosmology. Because of additional degree(s) of license seem to be significantly demanded, hence, the two 
accelerated phases cannot be undoubtedly expressed through the standard model of particle physics and general relativity.On another side, 
certain extra degrees can be assigned  for liberty to new, alien frames of matter, collectively termed as dark energy (DE)\cite{ref18,ref19,ref20}. 
The basic applicant for dynamical DE are phantom $(\omega <-1)$ \cite{ref21}, quintessence $(-1 < \omega < -1/3)$ \cite{ref22,ref23} (see \cite{ref24} 
for detailed review), quintom \cite{ref25} etc. Time, scale factor or redshift are the functions of the equation of state (EoS) of dynamical 
DE \cite{ref26,ref27}. On the other hand, from a transformed data of gravitation that adds general relativity (GR) as a low-energy absolute 
originates through the analysis of the gravitational origin \cite{ref28,ref29,ref30,ref31,ref32}.\\

To illustrate the prevailing accelerated enlargement of the cosmos, freshly, an exceptional concern has been held during the search of 
the holographic dark energy paradigm \cite{ref33,ref34,ref35,ref36,ref37,ref38,ref39,ref40,ref41,ref42,ref43,ref44}  in the light of holographic 
principle \cite{ref18,ref45,ref46,ref47} by determining and delineating the conventional holographic energy density as  
$\rho_{D}= 3c^{2}M^{2}_{pl}L^{-2}$,here, $c$ is a mathematical constant that depends on the entropy- area association of black holes \cite{ref33}.  
Latterly, a latest THDE model by transforming standard THDE as $S_{\delta}= \gamma A^{\delta}$, here $\gamma$ is an obscure constant and  
$\delta$ portrays the parameter of non- additivity, which adds including studied, called Tsallis holographic dark energy (THDE)\cite{ref48}, 
practicing Tsallis generalized entropy \cite{ref49}, with Hubble horizon in the light of the IR cutoff, in juncture through the thermodynamic 
analyses \cite{ref50,ref51}.\\

At the limit of $\gamma = 1/4G$ and $\delta = 1$ ( where $h = k_{B} = c = 1$ in units) Bekenstein entropy is redeemed. The method can be 
particularized by the conventional probability occurrence and the power-law allocation of probability display ineffective at this limit 
\cite{ref49}. In courses of effects in the cosmological and holographic situation, opens a surplus of possibilities 
\cite{ref50,ref51,ref52,ref53,ref54} and the quantum gravity too asserts this relation \cite{ref19}. It resembles the 
holographic postulate which repeats that the number of degrees of range of a physical system should be in a phalanx with bounding area, 
not with its capacity \cite{ref45,ref46} and this should be compelled by an IR cutoff. A similarity linked cutoffs with UV  
($\Lambda$), the IR ($L$) and the system entropy ($S$) is formed by Cohen et al. \cite{ref33}, $L^{3} \Lambda^{3}\leq S^{\frac{3}{4}}$, 
which after combining with $S_{\delta}= \gamma A^{\delta}$, leads to \cite{ref33} $\Lambda^{4} \leq (\gamma(4\pi)^{\delta})L^{2\delta-4}$, 
where $\Lambda^{4}$ renders the energy density of DE ($\rho_{D}$)  and the vacuum energy density, in the THDE formulas. The THDE as 
$\rho_{T} = CL^{2\delta-4}$ by the application of this inequality, where $C$ is an undefined parameter\cite{ref41,ref42,ref50}. The above 
expression gives the standard THDE with $C= 3c^{2}M_{p}^{2}$ and $c^{2}$, the modal parameter. The IR cutoff is a fit candidate to the 
Hubble horizon, so we have acknowledged a flat FRW universe.  In this state ($H^{-1} = L $). $L$ is the inevitable horizon applied near 
THDE, a harmonious formulation of THDE, is furnished, in \cite{ref55}. In non-interacting and interacting both the states, the dynamics of
the FRW flat universe is reviewed \cite{ref56,ref57,ref58}. Several authors \cite{ref58a,ref58b,ref58c,ref58d,ref58e} have recently investigated 
the different scenario of THDE models.
\\

The present and early stages of the universe were extensively considered as a good approach for the spatially isotropic and homogeneous, 
FRW model. The new experimental tests and logical arguments like Planks collaboration \cite{ref59}, Wilkinson Microwave Anisotropy Probe (WMAP) 
\cite{ref60,ref61} and Cosmic Background Explorers(COBE) \cite{ref62}, confirm the presence concerning an anisotropic state that resembles 
an isotropic unit. Consequently, the appearance of dark energy to follow the models of the universe with an anisotropic background makes a 
real thought. These best and mildest anisotropic ideals are Bianchi type models, which notwithstanding slightly and effectively express the 
anisotropic consequences. Though Bianchi's model universe is anisotropic yet by cosmological outlook, in an ancient epoch, the universe might 
seem anisotropic and also in the route of its progression, certain aspects might rub out on following the act of any manners or tools, appearing 
in the universe which is anisotropic and homogeneous.\\

In this research, we have interestingly examined the perfect cosmic model which is anisotropic and spatially homogeneous in the 
attendance of a winning bulky scalar field in the preparation of B-III space-time in $f(R, T)$ gravity. These anisotropic plus 
homogeneous nature of the Bianchi model plays a critical role and is very useful in the presence of the universe 
on a grand scale. In literature, various researchers have investigated these models by different aspects in detail 
\cite{ref63,ref64,ref65,ref66,ref67,ref68,ref69,ref70,ref71}. In the universe Bianchi type-III model loaded with the matter.  
New holographic dark energy in B- III universe including k-essence, HDE model of  B-III amidst quintessence and new HDE components 
are being reviewed \cite{ref72,ref73}. \\

Being motivated by these facts, we have read in this writing about  THDE (Tsallis holographic dark energy model) with time-dependent 
deceleration parameter under Bianchi-III model. The writing is served as: In Section $2$, the field equations and metric for the THDE 
model is exhibited. In Section $3$, we have found Model's physical properties and solutions. EoS parameter, deceleration parameter, and 
statefinder parameters are also defined in section $3$. Conformity among the scalar field model of quintessence and THDE Model are best 
illustrated in Section $4$. Section $5$ illustrates the distance cosmology- Luminosity distance, Angular-diameter distance and Distance 
modulus. In Section 6, we have presented the conclusion of our results.\\


\section{The Field Equations and Metric }
The  anisotropic and contiguous compatible Bianchi type-III metric defined as: 
\begin{equation}
\label{eq1}
ds^2 = dt^2 - A^2 dx^2 - e^{2mx} B^2 dy^2 - C^2 dz^2 ,
\end{equation}
where functions of cosmic time (t) are taken as $A$, $B$, and $C$. This specimen m  is a tyrannical positive constant in the space-time. 
Einstein's field equations of general relativity  (in gravitational entireties c~=~8$\pi$G~=~1) presented as to

\begin{equation}
\label{eq2}
R_{ij} - \frac{1}{2} Rg_{ij} = - (\bar{T}_{ij}+T_{ij
}),
\end{equation}
here the metric tensor is presented as $g_{ij}$, Ricci scalar is with R and the Ricci tensor is by $R_{ij} $.  $\bar{T}_{ij}$ including $T_{ij}$ are 
the energy-momentum tensors about THDE and matter, sequentially, and they are marked as
\begin{equation}
\label{eq3}
{T_{ij} = \rho_M u_i u_j} ~~ and ~~ {\bar{T}_{ij} = (\rho_T + p_T) u_i u_j - g_{ij} p_T} ,
\end{equation}
where  $\rho_M$ and $\rho_T$ are matter and THDE densities of energy, severally, and $p_T$ is the THDE pressure. The four-velocity 
vector $u^i$ is deemed to perform $u^iu_i$=1. Hither, we reflect the density of energy THDE in general relativity as:
\begin{equation}
 \label{eq4}
 \rho_T = \alpha H^{(-2\delta + 4)}.
\end{equation}
Here, $\alpha$ moreover $\delta$ are constants which need to provide the confinements inflicted by the recent observational data and 
$H$ is the Hubble parameter.\\ 

In a comoving coordinate policy, for Bianchi type-III space-time (\ref{eq1}), the Einstein's field equations (\ref{eq2}) with the help of 
Eq. (\ref{eq3}) finally begin to the next way of equalizations:
\begin{equation}
\label{eq5}
\frac{\ddot C}{C}+\frac{\ddot B}{B}+\frac{\dot B \dot C}{BC} = -\omega_T \rho_T,
\end{equation}
\begin{equation}
\label{eq6}
\frac{\ddot A}{A}+\frac{\dot A \dot C}{AC}+\frac{\ddot C}{C} = -\omega_T \rho_T,
\end{equation}
\begin{equation}
\label{7}
\frac{\ddot A}{A}+\frac{\dot A \dot B}{AB}+\frac{\ddot B}{B}-\frac{m^{2}}{A^{2}} =  -\omega_T \rho_T ,
\end{equation}

\begin{equation}
\label{eq8}
\frac{\dot A \dot B}{AB}+\frac{\dot A \dot C}{AC}+\frac{\dot C \dot B}{CB}-\frac{m^2}{A^2} = \rho_M + \rho_T,
\end{equation}
\begin{equation}
\label{eq9}
\frac{\dot B}{B}-\frac{\dot A}{A}=0,
\end{equation}
where $\omega_T = \frac{p_T}{\rho_T}$ is the equation of state (EoS) parameter of THDE and a hanging dot indicates differentiation 
concerning vast time t.\\

We obtain on integrating Eq. (\ref{eq9}), 
\begin{equation}
\label{eq10}
 B c{_1} = A,
\end{equation}
wherever ${c_1}$ is an integrating constant and, without any deterioration of generality, we take ${c_1}= 1$, so that we got\\
 \begin{equation}
\label{eq11}
     B = A.
     \end{equation}
The field Eqs. (\ref{eq5}) to (\ref{eq8}) will reduce in view of Eq. (\ref{eq11}), to 
\begin{equation}
\label{eq12}
\frac{\ddot C}{C}+\frac{\dot B \dot C}{BC}+\frac{\ddot B}{B} =- \omega_{T} \rho_{T}= -p_{T},
\end{equation}
\begin{equation}
\label{eq13}
\frac{\dot A \dot C}{AC}+\frac{\ddot A}{A}+\frac{\ddot C}{C} = -\omega_T \rho_T = -p_T,
\end{equation}
\begin{equation}
\label{eq14}
\frac{2\ddot A}{A}+\frac{\dot A^2}{A^2}-\frac{m^{2}}{A^{2}} = -p_T ,
\end{equation}
\begin{equation}
\label{eq15}
\frac{\dot A^2}{A^2}+\frac{2\dot A \dot C}{AC}-\frac{m^2}{A^2} = \rho_M + \rho_T.
\end{equation}
The equation of energy conservation $(T_{ij}+\bar{T_ij})_{;j} = 0$ can be achieved as
\begin{equation}
\label{eq16}
\dot{\rho_M}+\left(\frac{2\dot A}{A}+\frac{\dot C}{C}\right)\rho_M+\dot{\rho_T}+\left(\frac{2\dot A}{A}+\frac{\dot C}{C}\right)(1+\omega_T)\rho_T = 0.
\end{equation}
Here, an insignificant intercommunication is assumed by us between THDE and matter. Henceforth, they keep separately as
\begin{equation}
\label{eq17}
\dot{\rho_M}+\left(\frac{\dot C}{C}+\frac{2\dot A}{A}\right)\rho_M = 0 , 
\end{equation}
\begin{equation}
\label{eq18}
\dot{\rho_T}+\left(\frac{2\dot A}{A}+\frac{\dot C}{C}\right)(1+\omega_T)\rho_T = 0,
\end{equation}
On differentiate Eq. (\ref{eq4})
\begin{equation}
\label{eq19}
\rho_T = \alpha (-2\delta+4)\dot{H} H^{-2\delta+3}.
\end{equation}
Using Eqs. (\ref{eq4}) and (\ref{eq19}) in (\ref{eq18}),
\begin{equation}
\label{eq20}
\dot{\omega}_{T} = \frac{p_{T}}{\rho_{T}} = -1 + \frac{(2\delta-4)\dot{H}}{3 H^2}.
\end{equation}
Now, some essential assets of Bianchi type-III Universe are proffered by us, which is helpful to study the phylogeny of the Universe. 
The average scale factor ($a$), the volume ($V$), Hubble's parameter ($H$), the scalar expansion $(\theta$), the deceleration parameter ($q$), 
shear scalar $(\sigma^2)$, and anisotropy parameter $(\Delta)$ are given by  

\begin{equation}
\label{eq21}
a = (CAB)^ {1/3},
\end{equation}
\begin{equation}
\label{eq22}
V = a^3 = ABC,
\end{equation}
\begin{equation}
\label{eq23}
H = \frac{1}{3}\left(\frac{\dot A}{A}+\frac{\dot C}{C}+\frac{\dot B}{B}\right) = \frac{1}{3} \left(\frac{\dot C}{C}+\frac{2\dot A}{A}\right),
\end{equation}
\begin{equation}
\label{eq24}
\theta = \left(\frac{\dot B}{B}+\frac{\dot A}{A}+\frac{\dot C}{C}\right) = \left(\frac{\dot C}{C}+\frac{2\dot A}{A}\right),
\end{equation}
\begin{equation}
\label{eq25}
q = -\frac{a \ddot a}{\dot a^2} = -1-\frac{\dot H}{H^2},
\end{equation}
\begin{equation}
\label{eq26}
\sigma^2 = \frac{1}{2} \left(\sum_{i=1}^{3} H_i ^2 - \frac{\theta^2}{3}\right),
\end{equation}
\begin{equation}
\label{eq27}
\Delta = \frac{1}{3}\sum_{i=1}^{3}\left(\frac{\Delta H_i}{H}\right)^2,
\end{equation}
where  $H_i$ shows the directional Hubble parameters along with $x$, $y$ and $z$ directions and $\Delta H_i$ = $H_i - H$ (i=1,2,3) respectively.


\section{Fitting hybrid expansion law and solutions in our model }
We have five unknown parameters $C$, $A$, $\rho_T$, $p_T$ and $\rho_M$ including three independent field equations (\ref{eq12})-(\ref{eq15}). Brace 
supplementary confinements linking to certain parameters are needed to secure exact solutions of this scheme.\\

In beginning, we believe that in the model the expansion scalar $(\theta)$  is proportionate to shear scalar $(\sigma)$ \cite{ref74,ref75}, 
which addresses the correlation among the metric potentials as

\begin{equation}
\label{eq28}
C = A^n
\end{equation}

wherever $n \ne 1$  takes charge of the anisotropy of the Universe being a positive constant. \\

To provide exponent or power law, a constant deceleration parameter and therein references \cite{ref76}$-$\cite{ref79} has been practiced 
in the research. In the viewpoint of current observations of Planck Collaboration \cite{ref16}, SNIa (Type Ia supernova) \cite{ref5}$-$\cite{ref9}, 
and WMAP collaboration \cite{ref10,ref80,ref81}, the necessity of a time-dependent deceleration parameter is discussed in the introduction 
that defines expanded acceleration at present and expanded deceleration at past, hence there must be a shift from decelerated to acceleration 
phase in the Universe. In signature, the shift must be presented in the deceleration parameter  \cite{ref82}$-$\cite{ref84}.\\
 
A well-motivated ansatz first acknowledged through Abdussattar and Prajapati \cite{ref85} is viewed by us presently that restricts the 
functional structure of $q$ ( the deceleration parameter) as
 \begin{equation}
 \label{eq29}
 q=\frac{ n k}{(t+ k)^{2}}-1,
 \end{equation}
 here  $n>0$ (dimensionless) and $k>0$ (square of time's dimension) are constants. We discern $q=0$  while $t=\sqrt{k n}-k$ for such solution of 
 scale factor. For $t>\sqrt{k n}-k$, we obtain $q<0$ (i.e. accelerated expansion) and for $t<\sqrt{k n}-k$, we receive $q>0$ 
 (i.e. decelerated expansion). On integrating Eq. (\ref{eq25}), we obtain the scale factor as
 \begin{equation}
 \label{eq30}
 a(t)=c_2 \exp\int\frac{dt}{\int (q+1)dt+c_1},
 \end{equation}
 where $c_1, c_2$ are integrating constants.\\
 
 To obtain the scale factor, selecting suitable values of the constants ($c_{2}=1$ and $c_{1}=n$) with the help of  (\ref{eq29}), one can integrate  
 Eq. (\ref{eq30})
  \begin{equation}
  \label{eq31}
  a(t)=t^{\beta} \exp(\gamma t ),
  \end{equation}
  where $\beta > 0$ and $\gamma > 0$ are constants.\\

 With the modernization of scalar field of observational constraints, Akarsu {\it et al.} \cite{ref86} applied HEL(hybrid expansion law) 
 and the history of cosmology. HEL with integrating cosmic fluid was used by Avil$e'$s {\it et al.} \cite{ref87}. For resolving distinct cosmological 
 issues in the theory of GR and $f(R, T)$ gravity, numerous authors have studied HEL  \cite{ref88}$-$\cite{ref94a}. Moraes {\it et al.} \cite{ref96} 
 and Moraes \cite{ref95}, also did some research. Lately, by practicing HEL Moraes and Sahoo \cite{ref97} reviewed the $f(R, T)$ gravity in 
 non-minimal matter geometry coupling. \\

 By Eqs. (\ref{eq11}), (\ref{eq22}), (\ref{eq27}), and (\ref{eq31}) we get
\begin{equation}
\label{eq32}
B = A = (e^{\gamma t} t^{\beta})^\frac{3}{n+2}
\end{equation} 
\begin{equation}
\label{eq33}
C =  (e^{\gamma t} t^{\beta})^\frac{3n}{n+2} 
\end{equation}
Hence, with the use of Eqs. (\ref{eq32}) and (\ref{eq33}) the model (\ref{eq1}) is reduced to
\begin{equation}
\label{eq34}
ds^2 = dt^2 - (e^{\gamma t} t^{\beta})^\frac{6}{n+2} [dx^2 - e^{2mx} dy^2] - (e^{\gamma t} t^{\beta})^\frac{6n}{n+2} dz^2.
\end{equation}
On differentiating Eqs. (\ref{eq32}) and (\ref{eq33}), we obtain
\begin{equation}
\label{eq35}
\dot A = \dot B = \frac{3}{n+2}\left(\gamma + \frac{\beta}{t}\right) \left(e^{\gamma t} t^{\beta}\right)^\frac{3}{n+2}
\end{equation}
\begin{equation}
\label{eq36}
 \dot C = \frac{3n}{n+2}\left( \frac{\beta}{t}\gamma+\gamma \right) \left(e^{\gamma t} t^{\beta}\right)^\frac{3n}{n+2}
\end{equation}

From Eqs. (\ref{eq32}), (\ref{eq33}), (\ref{eq35}) and (\ref{eq36}), we obtain
\begin{equation}
\label{eq37}
H_x = \frac{\dot A}{A} = \frac{3}{n+2}\left(\gamma+\frac{\beta}{t}\right),
\end{equation}
\begin{equation}
\label{eq38}
H_y = \frac{\dot B}{B} = \frac{3}{n+2}\left(\gamma+\frac{\beta}{t}\right),
\end{equation}
\begin{equation}
\label{eq39}
H_z = \frac{\dot C}{C} = \frac{3n}{n+2}\left(\gamma+\frac{\beta}{t}\right),
\end{equation}
\begin{equation}
\label{eq40}
H = \gamma + \frac{\beta}{t}.
\end{equation}

For the present universe $H_{0} = 0.73$ with $q_{0} = -0.54$ (Giostri {\it et al.} \cite{ref98}), Eqs. (\ref{eq40}) and (\ref{eq54}) help to find the 
relation between the constants $\beta$ and $\gamma$:
\begin{equation}
\label{eq41}
2505.3624 ~\beta =5446.44 + \gamma^{2} - 147.6~ \gamma
\end{equation}
On putting $\gamma = 1$ and $0.5$ in Eq. (\ref{eq41}), we get $\beta = 2.1154$, and $2.1445$ respectively.\\

 Eqs. (\ref{eq24}) and (\ref{eq40}) reduce to
\begin{equation}
\label{eq42}
\theta = 3 H = 3\left(\gamma + \frac{\beta}{t}\right)
\end{equation}
Using Eqs. (\ref{eq37}), (\ref{eq38}), and (\ref{eq30}) in (\ref{eq26}) and (\ref{eq27}), we obtain
\begin{equation}
\label{eq43}
\sigma^2 = 3\left(\frac{n-1}{n+2}\right)^2 \left(\gamma + \frac{\beta}{t}\right)^2
\end{equation}
\begin{equation}
\label{eq44}
\Delta = 2 \left(\frac{n-1}{n+2}\right)^2
\end{equation}
Differentiate Eq. (\ref{eq40}), we get
\begin{equation}
\label{eq45}
\dot H = - \frac{\beta}{t^2}
\end{equation}
From Eqs. (\ref{eq22}) and (\ref{eq31}) we get
\begin{equation}
\label{eq46} 
V =  t^{3\beta}  e^{3\gamma t}
\end{equation}
For our model, the redshift parameter ($z$) can be written as follows:
\begin{equation}
\label{eq47}
z = \frac{a_0}{ t^\beta e^{\gamma t}} - 1
\end{equation}
Here at z=0,  $a_0$ is the existing value of the scalar factor.
From Eqs. (\ref{eq42})-(\ref{eq46}), we perceive that at the primary period the spatial volume V is zero  (i.e. at $t=0$)  
and the additional parameters $\sigma$, $H$, and $\theta$ deviate at this era,  $\theta$, $\sigma$, and $H$ all tend to zero if  
t$\to$ $\infty$, spatial volume V$\to$ $\infty$. Hence the model springs evolving at zero volume with an infinite rate of expansion 
and through the development of model this expansion pace slows down. As $\Delta$ = constant $\neq$ 0 and the isotropy state 
$\frac{\sigma}{\theta}$ = consistent $\neq$ 0 for $n \neq 1$, in the evolution the model is anisotropic.\\

Using Eqs. (\ref{eq4}) and (\ref{eq40}), we found
\begin{equation}
\label{eq48}
\rho_T = \alpha \left( \gamma + \frac{\beta}{t} \right)^{-2\delta+4}
\end{equation}
Using Eqs. (\ref{eq35}), (\ref{eq36}) in (\ref{eq17}) and then by integrating, we obtain
\begin{equation}
\label{eq49}
\rho_M = \frac{d_1}{t^{3\beta} e^{3\gamma t}}
\end{equation}
where $d_1 \neq 0$ is a constant of integration. \\

\begin{figure}[htbp]
	\centering
	\includegraphics[width=7cm,height=7cm,angle=0]{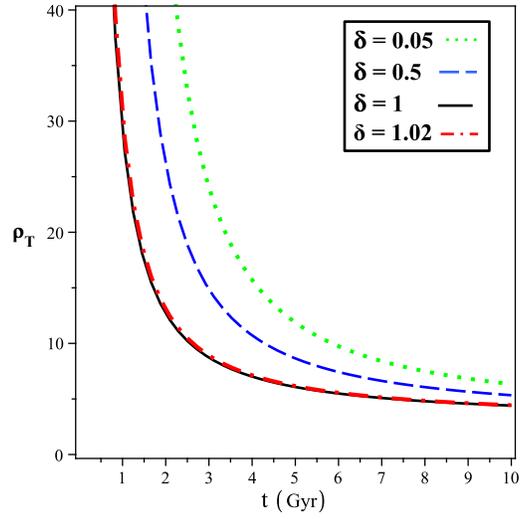}
	\caption{Plot of energy density of THDE $(\rho_{T})$ at variance with cosmic time ($t$) for $\alpha = 3$, $\beta = 2.1154$, 
	$\gamma = 1$, $\delta = 0.05,0.5,1 ~ and~1.02 $} 
\end{figure}
\begin{figure}[htbp]
	\centering
	\includegraphics[width=7cm,height=7cm,angle=0]{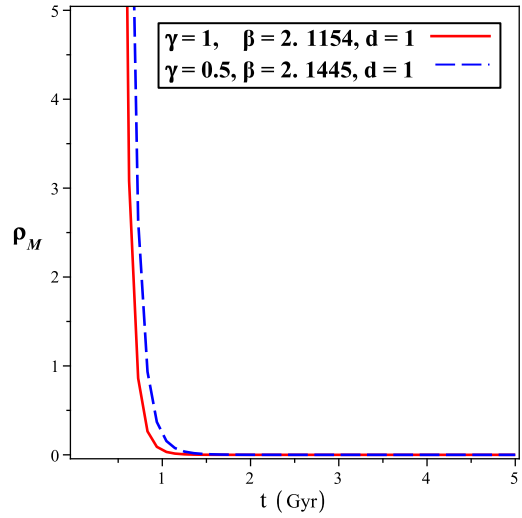}
	\caption{Plot of energy density of matter $(\rho_{M})$ at variance with cosmic time ($t$) for $\beta = 2.1154, 2.1445$, 
	$\gamma = 1, 0.5$, $d_{1}=1$} 
\end{figure}

From Eqs. (\ref{eq49}) and (\ref{eq48}), we discern that the matter's energy density is a limiting function of $t$, while the energy density 
of THDE is a decreasing or increasing function depend upon $\delta$. We have framed them against time, as shown in Figs. $1$ and $2$, to 
explain the definite nature of $\rho_M$ and $\rho_T$. Hereabouts, we contemplate the arbitrary constants $\alpha = 3$, $\beta = 2.1154$ and 
$\gamma = 1$. It can be noticed that $\rho_T$ is decreasing for all the values of $\delta$ $(\delta < 2)$, while $\rho_M$ is clipping through 
the progression of a model for $\beta$ and $\gamma$. Interestingly, we have noticed that the THDE's energy density does not fade, whereas 
the energy density of matter disappears during the evolution for enough long values of the time. The dynamic results of this analysis intimate 
that the lowering of the energy density of THDE worrying for $t$ traverses the volume extension of the Universe.\\

\begin{figure}[htbp]
	\centering
	\includegraphics[width=7cm,height=7cm,angle=0]{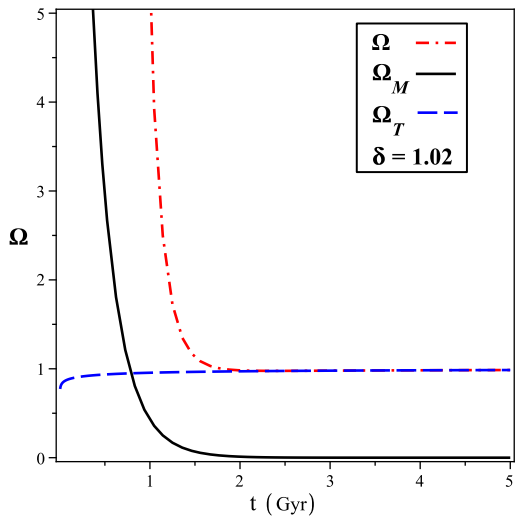}
	\caption{The plot of energy density parameter $\Omega$ concerning cosmic time ($t$) for $\alpha = 3$, $\beta = 2.1154$, $\gamma = 1$, 
	$\delta = 1.02 $} 
\end{figure}

The matter density ($\Omega_M$) and THDE density ($\Omega_T$) parameter are presented severally, by 
\begin{equation}
\label{eq50}
\Omega_M = \frac{\rho_M}{3 H^2} = \frac{d_1}{3 e^{3\gamma t} t^{3\beta}\left(\gamma + \frac{\beta}{t}\right)^{2}}
\end{equation}
\begin{equation}
\label{eq51}
\Omega_T = \frac{\rho_T}{3 H^2} = \frac{\alpha}{3}\left(\gamma + \frac{\beta}{t}\right)^{-2\delta+2}
\end{equation}

We find overall density parameter by Eqs. (\ref{eq50}) and (\ref{eq51}),
\begin{equation}
\label{eq52}
\Omega = \Omega_M + \Omega_T = \frac{d_1}{3 e^{3\gamma t} t^{3\beta}\left(\gamma + \frac{\beta}{t}\right)^2} + \frac{\alpha}{3}
\left(\gamma + \frac{\beta}{t}\right)^-2\delta+2
\end{equation}
  
 Figure $3$ illustrates the nature of overall density parameter $(\Omega)$, density parameter of THDE $(\Omega_{T})$ and matter 
$(\Omega_{M})$  at variance with cosmic time $t$. It is presented in Fig. $3$ that for late times overall density parameter $(\Omega)$ 
approaches $1$. Accordingly, our THDE model prophesies an adequately large time that the anisotropy will damp out, and the Universe will 
display isotropic. The outcomes received are considerably related to the outcome got by Samanta and Mishra \cite{ref99}, where they obtained 
in their text that the Universe resembles isotropy for an amply wide time. Additionally, during $\delta = 1.02$, the density parameter shows 
an almost similar style to this event. 
\subsection{EoS parameter}

\begin{figure}[htbp]
	\centering
	4(a)\includegraphics[width=7cm,height=7cm,angle=0]{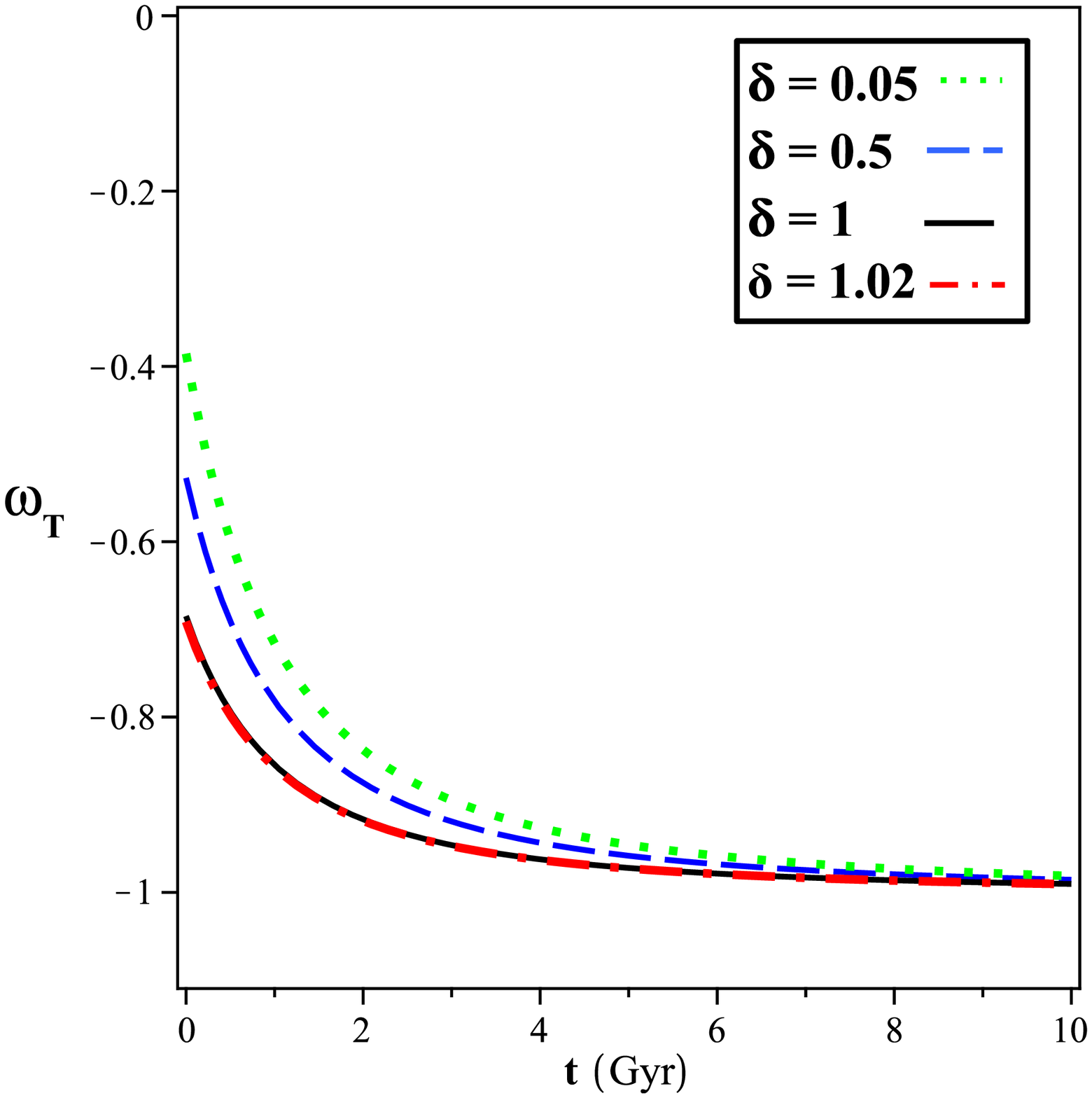}~~~
	4(b)\includegraphics[width=7cm,height=7cm,angle=0]{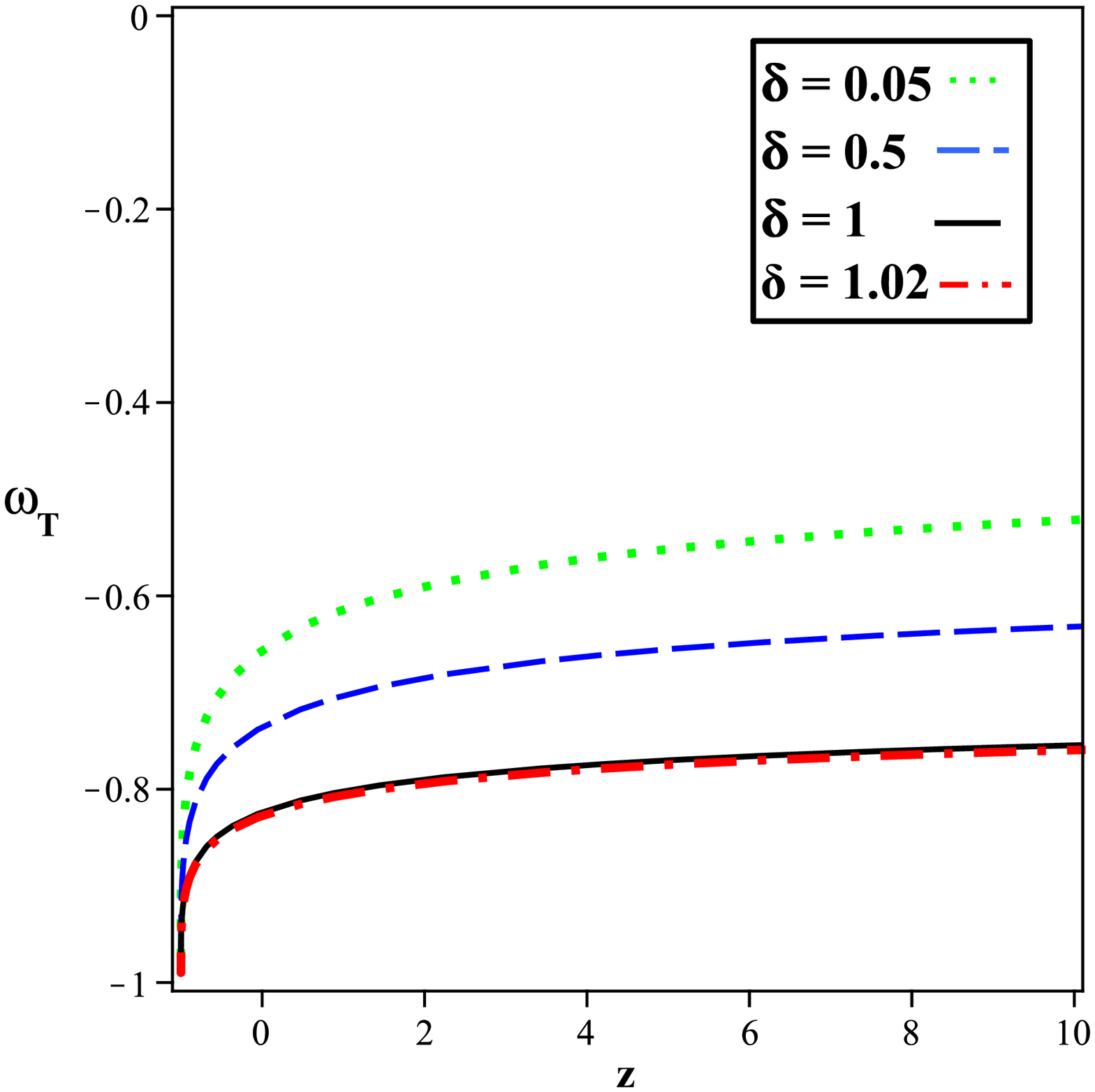}
	\caption{Plot of EoS $(\omega_{T})$ for cosmic time ($t$) and redshift ($z$) respectively, when $\beta = 2.1154$, $\gamma = 1$, 
	$\delta = 0.05, 0.5, 1$~ and ~$1.02$} 
\end{figure}
Using Eqs. (\ref{eq40}), (\ref{eq45}) in (\ref{eq30}), the Eos parameter of THDE is given by 
\begin{equation}
\label{eq53}
\omega_T = -1 - \frac{\beta(2\delta - 4)}{3(\gamma t + \beta)^2}
\end{equation} 
Eq. (\ref{eq53}) shows the EoS parameter of the model. From Fig. $4$, it can be seen that the EoS parameter of THDE model is time-dependent 
and converges to  $\omega_T \to -1$ with an increase in time. The figure displays the progression of the THDE EoS parameter $\omega_T$ at 
variance with infinite time $t$. This can be perceived that $\omega_T$ of the model diversifies in the quintessence field (-1$< \omega_T < -1/3$) 
during its evolving nature for all three different values of $\delta$. Furthermore, it can be seen that the EoS parameter resembles the 
$\Lambda$CDM model ($\omega_T = -1$) in eternity. This intimates that at more miniature charges of cosmic time $t$, the form has a more 
prominent accelerating impact.
\subsection{Deceleration parameter}
 In this model, the deceleration parameter is received by applying Eqs. (\ref{eq40}) and (\ref{eq45}) in Eq. (\ref{eq25}), we get 
\begin{equation} 
\label{eq54}
q = -1 + \frac{\beta}{(\gamma t + \beta)^2}
\end{equation}
this works as a sign of this presence of buildup of the model. The model decelerates conventionally if q$>0$, while the $q<0$ model 
designates inflation. From (\ref{eq54}), we witness that $q>0$ toward t$<\frac{\sqrt{\beta} - \beta}{\gamma}$ and $q<0$ for 
t$>\frac{\sqrt{\beta} - \beta}{\gamma}$. In this span of $-1\leq q <0$, the value of DP lies and, the Universe is accelerating these 
observations is exposed by SNe Ia recently. Fig. 5 portrays the DP at variance with $t$ for different alternatives of $\beta and \gamma$ 
in such a way that in recent researches, DP is in good compliance. Fig. $5$ points out that q reduces from positive to negative zone plus 
eventually points to $-1$. To read the nature of DP at variance with cosmic time $t$, we have framed $q$ in terms of $t$ in Fig. $5a$. 
For all the preferences of parameter $\gamma$,~$\beta$, the model flaunts a shift from the decelerated phase to an accelerated epoch. 
Earlier, the model enters into an accelerated phase for the bigger values of $\delta$.\\

Accordingly, our model is occurring to the modern accelerating scenario from an initial decelerating phase of the universe and from 
various experiments the observational data is consistent for the values of DP \cite{ref5,ref6}.

\begin{figure}[htbp]
	\centering
	5(a)\includegraphics[width=7cm,height=7cm,angle=0]{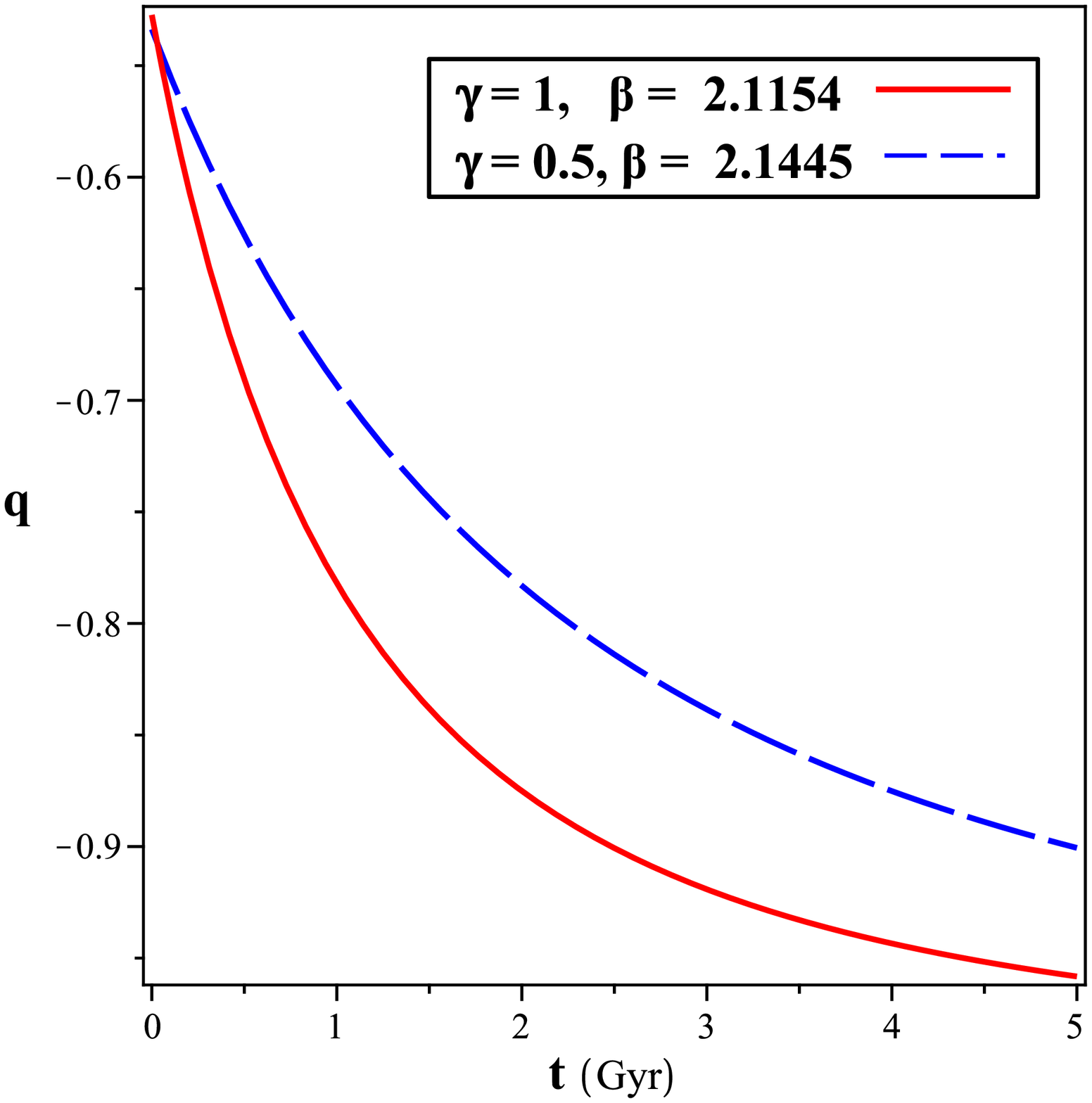}~~~
	5(b)\includegraphics[width=7cm,height=7cm,angle=0]{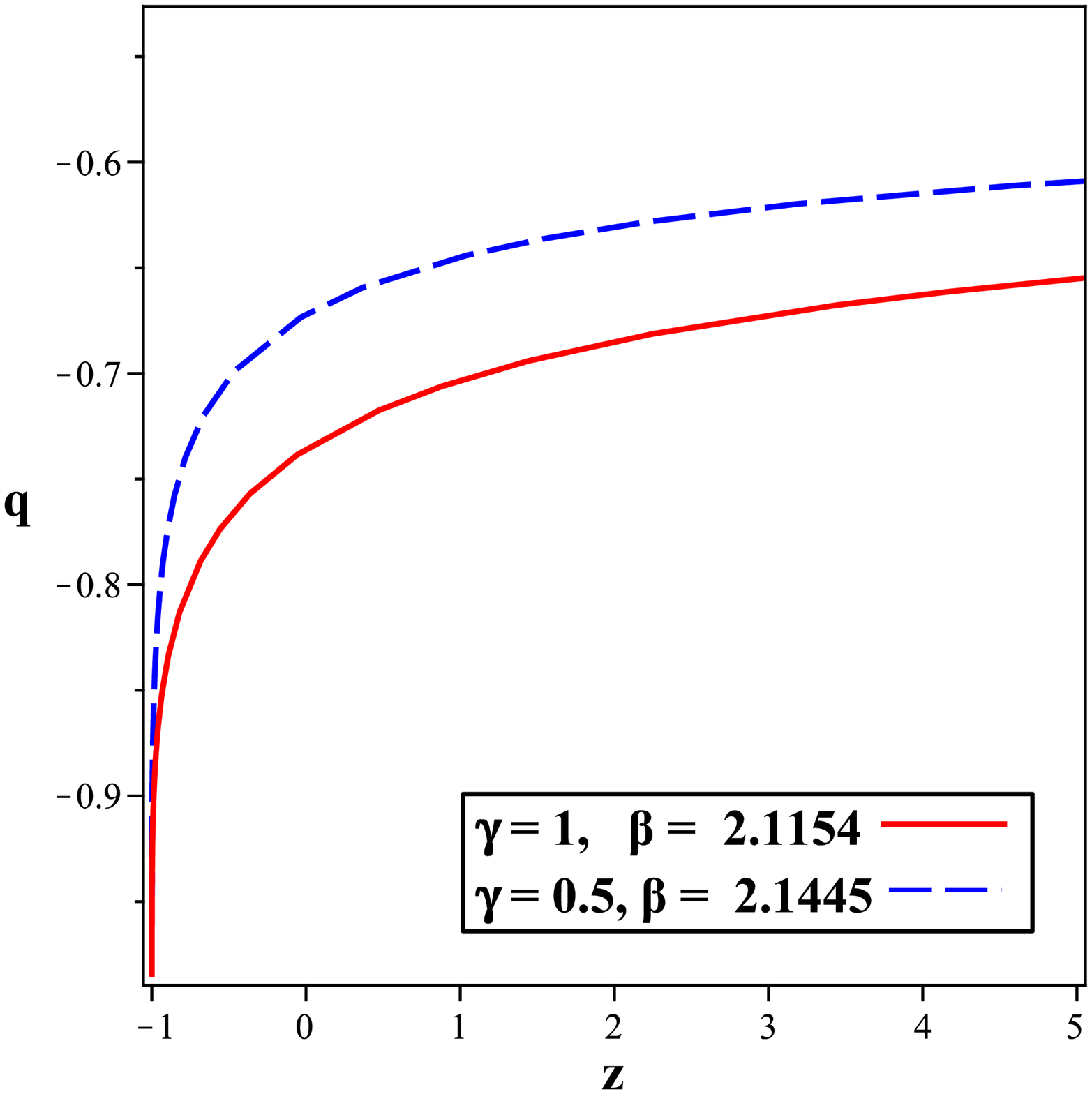}
	5(c)\includegraphics[width=7cm,height=7cm,angle=0]{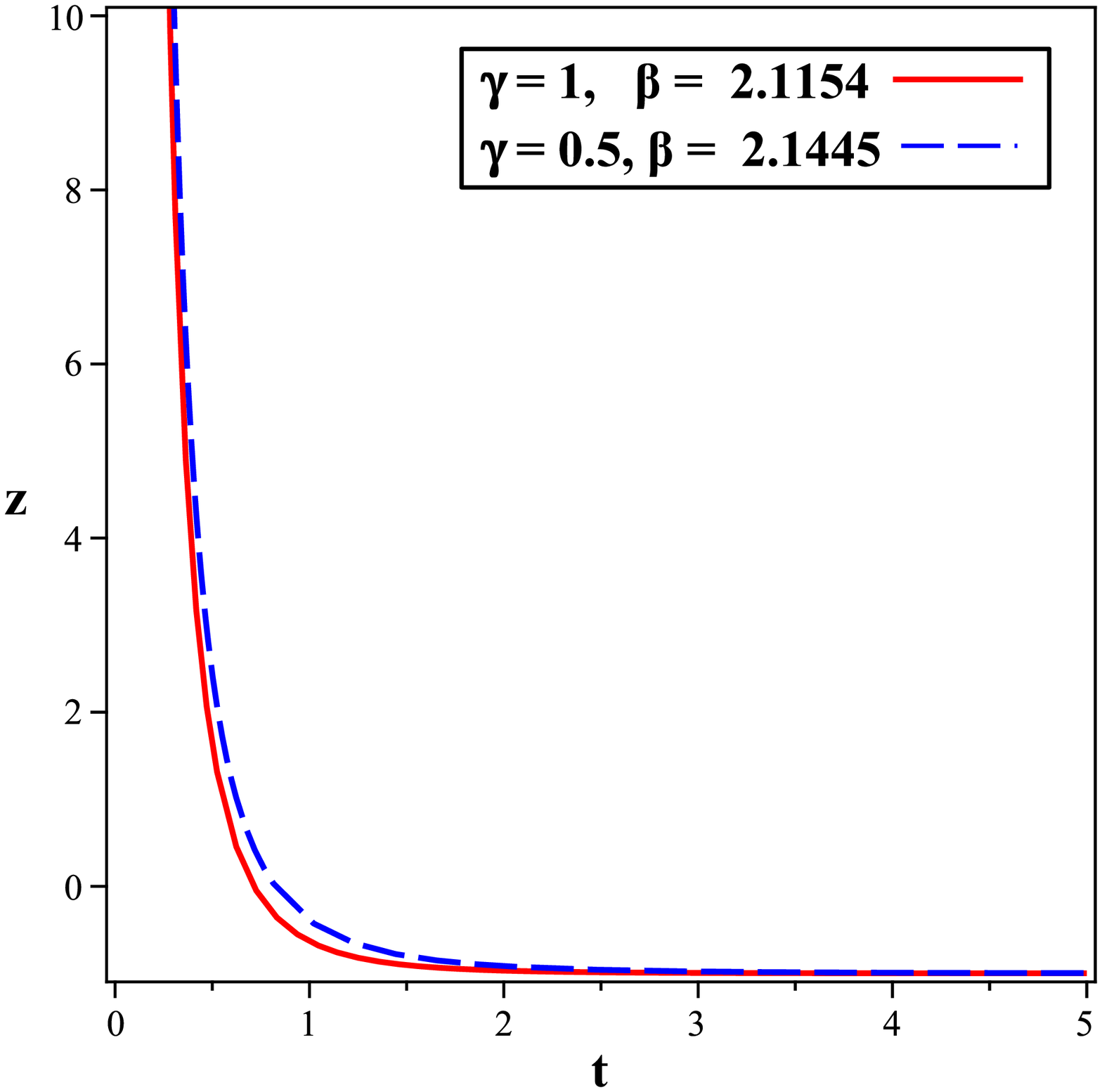}~~~
	\caption{5(a), (b) \& (c) are plots of $q$ versus  $t$, $q$ versus $z$ and $t$ versus $z$ respectively  
	for $\beta = 2.1154,2.1445$, $\gamma = 1,0.5$} 
\end{figure}

\subsection{Statefinder parameters}
To paraphrase or catch the cosmic acceleration many more DE models are being erected. To discriminate amid these competing cosmological 
synopses implicating DE, a fine-tuned and sturdy diagnostic for DE figures is a necessity. The nominal "statefinder" was founded by 
Sahni {\it et al.} \cite{ref101}. To fulfill one goal of a diagnostic scheme that presents usage of parameters pair {r,s}. Statefinder examines 
the increased dynamics of the Universe by higher derivatives of the increase factor and is an actual next level behind \textit{H} and $q$. 
The statefinder diagnostic is a beneficial way of identifying DE models, considering various cosmological models requiring DE to exhibit 
qualitative distinct evolutionary plots in the r-s plane. The parameters of statefinder resemble to {$r=1$, $s=1$} 
to the spatially flat TCDM model. The statefinder duo {r,s} can arise as:

\begin{equation}
\label{eq55}
r = \frac{\dddot a }{a h^3}, s = \frac{r-1}{3(q-\frac{1}{2})}
\end{equation}   
On differentiate Eq. (\ref{eq31}) thrice, we get
\begin{equation}
\label{eq56}
\dddot a = 3 \gamma^2 \beta e^{\gamma t} t^{\beta-1} + 3\gamma\beta(\beta-1) e^{\gamma t} t^{\beta-2} + \beta(\beta-1)
(\beta-2)e^{\gamma t} t^{\beta-3} + \gamma^3 e^{\gamma t} t^\beta
\end{equation}
Using Eqs. (\ref{eq31}), (\ref{eq40}) and (\ref{eq56}) in (\ref{eq55}), we found
\begin{equation}
\label{eq57}
r = \frac{\frac{3 \gamma^2 \beta}{t} + \frac{3\gamma \beta^2}{t^2} + \frac{\beta^3}{t^3} - \frac{3\gamma\beta}{t^2} - \frac{3\beta^2}{t^3} + 
\frac{2\beta}{t^3} + \gamma^3}{(\gamma + \frac{\beta}{t})^3}
\end{equation}
Using Eqs. (\ref{eq54}) and (\ref{eq57}) in (\ref{eq55}), we found
\begin{equation}
\label{eq58}
s = \frac{\frac{-3 \gamma \beta}{t^2} - \frac{3{\beta}^2}{t^3} + \frac{2\beta}{t^3}}{3(\gamma + \frac{\beta}{t})^{3}\left[\frac{-3}{2} + 
\frac{\beta}{\left(\gamma t + \beta\right)^2}\right]}
\end{equation}
From Eq. (\ref{eq58}), The contents of statefinder set ultimately based on the states of $t$ and fixed $\beta$ and $\gamma$ that 
is very fascinating to remark. It is known that cosmic time ($t$) persists to infinity when the statefinder pair presents $r=1$ and $s=0$. 
This reinforced that our THDE model would accord beside the flat $\Lambda$CDM rule in eternity.

\section{Conformity Among the Scalar Field Model of Quintessence and THDE Model}
With negative pressure, Quintessence is a dynamical, emerging and spatially inhomogeneous component. The key ideas to examine 
quintessence as one of the nominees of DE are as follows:
(i)  Shows elementary physics for different connotations.
(ii)  This reveals the cosmic event query
(iii) That suits the observational data strongly than the cosmological constant
(iv)  the innovative picture of the overall memoir of the universe is recommended.\\
 Unlike the quintessential pressure, energy density and a cosmological constant evolve with time, followed by the EoS parameter. 
 The energy density of quintessence shows a general principle connected with a scalar field $\phi$ leisurely twirling down a potential 
 $V(\phi)$. In the quintessence principles of DE, the potential energy of the dynamical field has discovered the acceleration in the 
 scale factor led to as the quintessence field.\\
 
For the quintessence scalar field model, the pressure and energy density is supplied  
\begin{equation}
  \label{eq59}
  \rho_{de} = \frac{\dot{\phi}^2}{2} + V(\phi);  ~~  p_{de} = \frac{\dot{\phi}^2}{2} - V(\phi) 
    \end{equation}
 which gives
 \begin{equation}
\label{eq60}
\dot{\phi}^2 = \rho_{de} + p_{de};~~ V(\phi) = \frac{\rho_{de} - p_{de}}{2} =\frac{ (1-\omega_{de})\rho_{de}}{2}
\end{equation}
By taking $\rho_{de}$=$\rho_T$ and $p_{de}$=$p_T$,we get
\begin{equation}
\label{eq61}
\phi = \sqrt{\frac{\alpha \beta (4-2\delta)}{3}}\int{\frac{(\gamma t +\beta)^{-\delta+1}}{t^{-\delta+2}}}
\end{equation}
We perceive the potential as
\begin{equation}
\label{eq62}
V(\phi)=\alpha\left(\gamma+\frac{\beta}{t}\right)^{-2\delta+4}\left[1+\frac{\beta(\delta-2)}{3(\gamma t+\beta)^2}\right]
\end{equation}
The quintessence parameter model's EoS parameter
\begin{equation}
\label{eq63}
\omega_{\phi}=\frac{p_T}{\rho_T}=\frac{\dot{\phi}^2-2V{(\phi)}}{\dot{\phi}^2+2V{(\phi)}}=-1-\frac{(2\delta-4)\beta}{3(\gamma t + \beta)^2}
\end{equation}
\begin{figure}[htbp]
	\centering
	6(a)\includegraphics[width=7cm,height=7cm,angle=0]{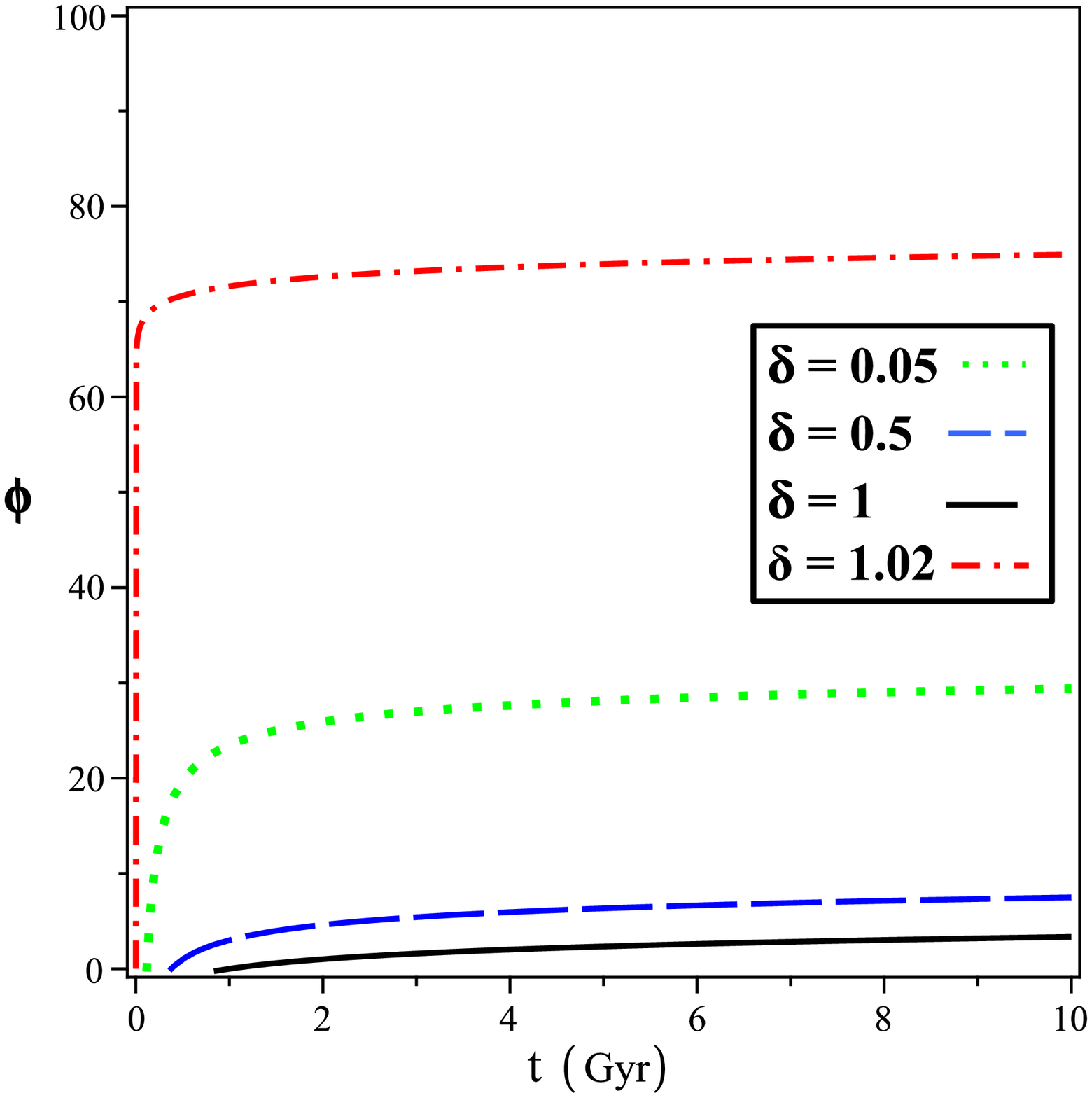}
	6(b)\includegraphics[width=7cm,height=7cm,angle=0]{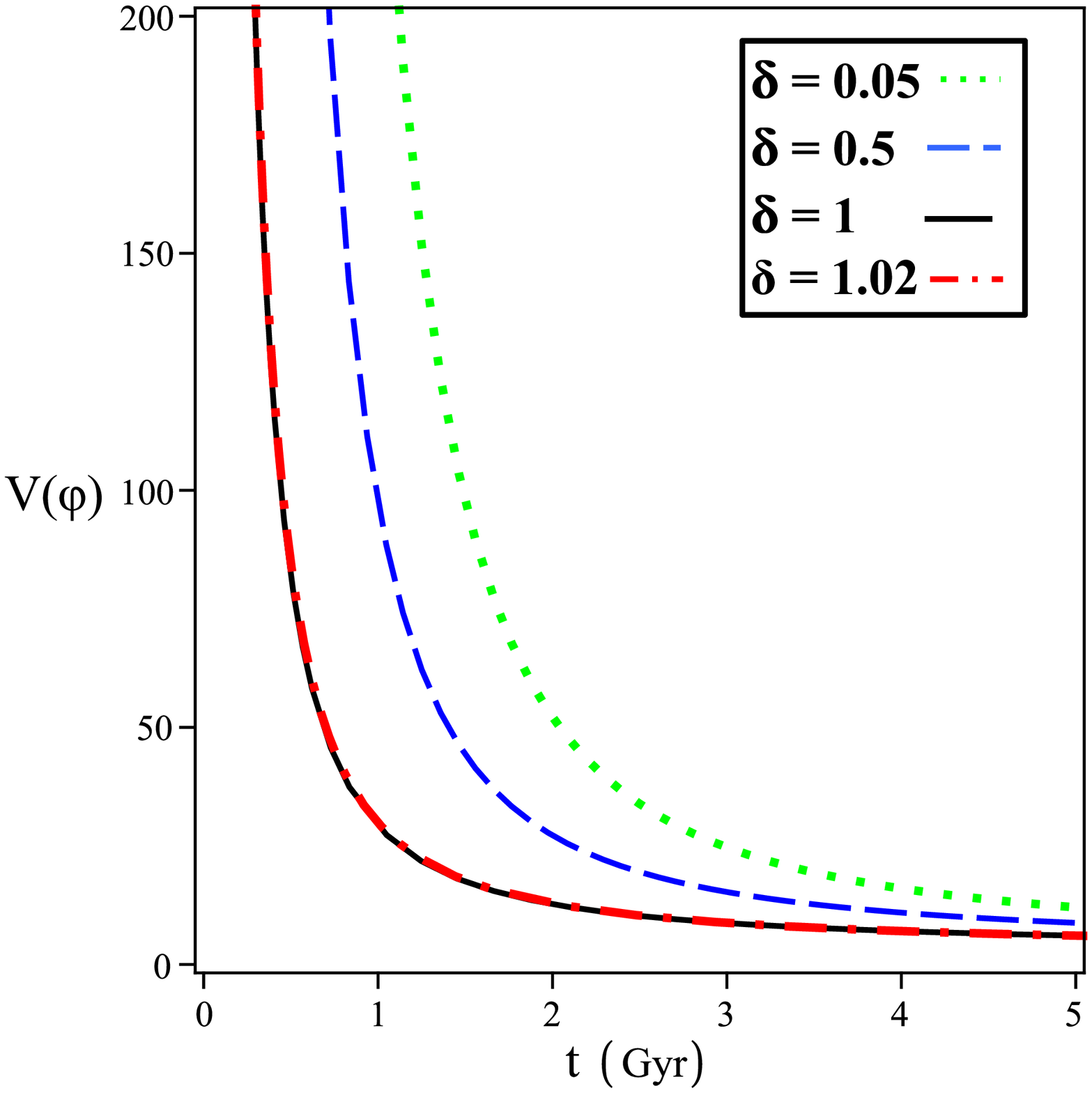}
	\caption{Plots of scalar field and potential energy for cosmic time $t$ respectively, when $\alpha = 3$, $\beta = 2.1154$, 
	$\gamma = 1$, $\delta = 0.05,0.5,1$ ~and~$1.02 $} 
 \end{figure}
The equalization of state parameter for quintessence must be more invisible than$\frac{-1}{3}$. For the accelerated expansion of the universe, 
the scalar field $\phi$ and potential $V(\phi)$ of quintessence scalar field model that writes to the B-III THDE model is practiced. 
Of Eq. (\ref{eq61}), the scalar field $\phi$ dissolves for liberal values of the time and displays a decreasing function of $t$. 
It recognizes from Eq. (\ref{eq62}) that the potential of scalar field $V(\phi)$ converges to $2$ as t$\to$ $\infty$ and 
demonstrates decreasing function of time.
\section{Distances in Cosmology}

One of the most fundamental measurements to work is named Distance. Sometimes distance measurement performs remarkably and an essential 
part to know about the Universe in the records of cosmology. We have displayed some of the modified distance measures in this section.
We describe the following observational quantities. Such type of studies have recently performed in \cite{ref101,ref102}.
\subsection{Luminosity distance}

Redshift-luminosity distance similarity is the most significant observational device (Carroll et al. \cite{ref103}, Liddle and Lyth \cite{ref104})to 
analyze the evolution of the Universe. We obtain the expression for luminosity distance $(d_{L})$ in variance with redshift, while the light 
appearing out from a distant luminous body gets redshifted because of the Universe's expansion. With the help of luminosity distance, we restrict 
the flux of a source. The aforementioned information is being practiced from \cite{ref105,ref105}, that is presented as
\begin{eqnarray}
\label{eq64}
d_{L} = r~ a_{0}~ (1+z),
\end{eqnarray}
here $r$ is the radial coordinate of the origin. Initially, we examine a ray of light having a radial coordinate. Therefore,
\begin{eqnarray}
\label{eq65}
r = \int_{0}^{r} {dr} = \int_{0}^{r}\frac{c~ dt}{a(t)} = \frac{1}{a_{0}} H_{0}\int_{0}^{z}\frac{c~dz}{h(z)}.
\end{eqnarray}
Here we have applied $dt = \frac{dz}{\dot{z}}$,
\begin{eqnarray}
\label{eq66}
\dot{z} = -H (z+1) ~~~~and~~~~ h(z) = \frac{H}{H_{0}}.
\end{eqnarray}
$H_{0}$ is Hubble parameter's current value and hence we got
\begin{eqnarray}
\label{eq67}
d_{L} = \frac{c~(1+z)}{H_0} \int_{0}^{z}\frac{dz}{h(z)}.
\end{eqnarray}
By using this expression we got luminosity as;
\begin{eqnarray}
\label{eq68}
d_{L} = a_{0}~ t^{-\beta} e^{-\gamma t} \int_{0}^{t} c~t^{-\beta} e^{-\gamma t} dt.
\end{eqnarray}
\begin{figure}[htbp]
	\centering
	\includegraphics[width=7cm,height=7cm,angle=0]{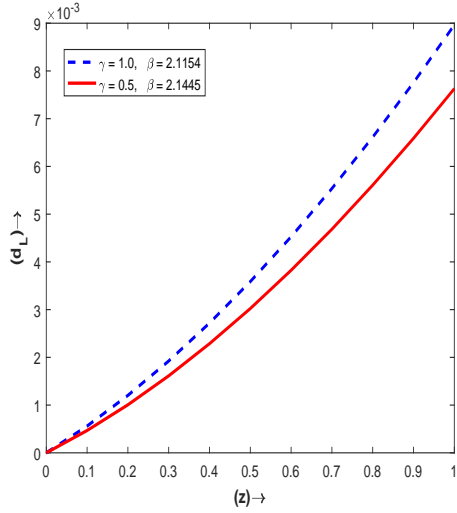}
	\caption{Variation of Luminosity distance ($d_L$) versus redshift z  for $\beta = 2.1154,2.1445$, $\gamma = 1,0.5$} 
\end{figure}
Figure 7 demonstrates the luminosity distance versus redshift $z$ for observational values of Type Ia supernovae (SNIa) data in combination 
with CMB and BAO observations of ($\beta, \gamma$). Here we find the luminosity distance $d_{L}$ as an increasing function of redshift.
\subsection{Angular-diameter distance}

The angular-diameter distance $d_{A}$ is termed as\\
$~~~~~~~~~~~~~~~~~~~~~~~~~~~~~~~~~~~~~~~~~~~~~ $ $\theta = \frac{l}{d_{A}}$,\\
where $\theta$ is the angle transverse by an object of size l. It is also defined in term of $d(z)$ and Luminosity distance as\\
$~~~~~~~~~~~~~~~~~~~~~~~~~~~~~~$ $d_{A} = (z+1)^{-1} d(z) = d_{L} (z+1)^{-2}$.\\
For the present model angular-diameter distance is as follows:
\begin{eqnarray}
\label{69}
d_{A} = a_{0}~ t^{\beta} e^{\gamma t} \int_{0}^{t} c~t^{-\beta} e^{-\gamma t} dt .
\end{eqnarray}
\begin{figure}[htbp]
	\centering
	\includegraphics[width=7cm,height=7cm,angle=0]{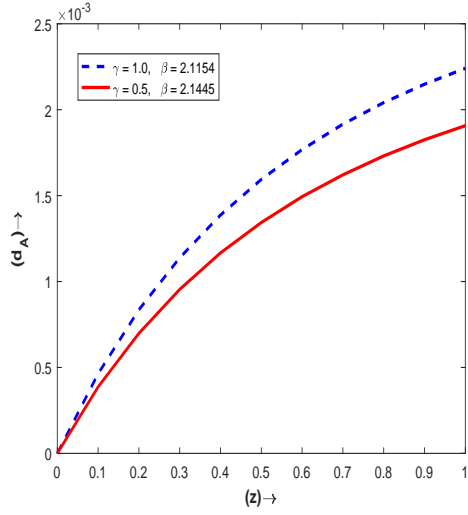}
	\caption{Plot of Angular distance versus redshift $z$ for $\beta = 2.1154,2.1445$, $\gamma = 1,0.5$} 
\end{figure}
We have plotted the variation of Angular diameter distance versus redshift $z$ for observational values of Type Ia supernovae (SNIa) data in combination 
with CMB and BAO observations of ($\beta, \gamma$) in Fig. $8$. Here we observe that the Angular diameter distance $d_{A}$ increases with the increase of 
redshift.
\subsection{Distance modulus}

The distance modulus $(\mu(z))$ can be written as\\

$~~~~~~~~~~~~~~~~~~~~~~~~~~~~~$ $(\mu(z)) = 5\log_{10}(d_{L}) + 25$. \\

In variance with redshift parameter ($z$), the distance modulus $(\mu(z))$ can be received as
\begin{eqnarray}
\label{eq70}
\mu(z) = 25+5~\log_{10}\Bigg( a_{0}~ t^{-\beta} e^{-\gamma t} \int_{0}^{t} c~t^{-\beta} e^{-\gamma t} dt\Bigg).
\end{eqnarray}
\begin{figure}[htbp]
	\centering
	\includegraphics[width=7cm,height=7cm,angle=0]{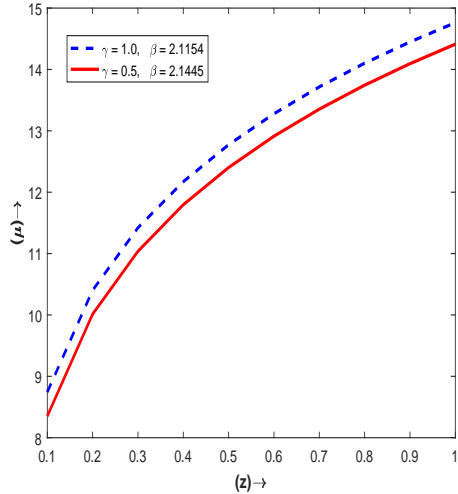}
	\caption{Plot of Distance modulus versus redshift z for $\beta = 2.1154,2.1445$, $\gamma = 1,0.5$} 
\end{figure}
We depict the variation of Distance modulus versus redshift $z$ for observational values of Type Ia supernovae (SNIa) data in combination 
with CMB and BAO observations of ($\beta, \gamma$) in Fig. $9$. Here we observe that the Distance modulus $\mu$ is also an increasing function of 
redshift.
\section{Conclusions}
In this research, we investigated the nature of the Tsallis holographic dark energy model (THDE) applying Bianchi type-III Universe with 
Hubble horizon as IR cut-off. We used two conditions in this research to obtain a deterministic explanation that the expansion scalar ($\theta$) 
is proportionate to the shear scalar ($\sigma$) and a hybrid expansion law for the scale factor $a = t^\beta e^{\gamma t}$, where 
$\beta>0$, $\gamma>0$. Our research is based on SNIa data in combination with CMB and BAO observations (Giostri et al, JCAP 3, 27 (2012), 
arXiv:1203.3213[astro-ph.CO]), the prevailing values of Hubble constant plus deceleration parameter are $H_{0} = 73.8$ and $q_{0} = -0.54$ 
respectively. Compiling our theoretical models with this data, we obtain $\beta = 2.1445~ \& ~ 2.1154$ for $\gamma = 0.5 ~ \& ~ 1$ respectively. 
In this research, we acquire numerous cosmological parameters and examined their evolution for a better description of the Universe's accelerated 
expansion.\\

We can see from Fig. 1 that $\rho_T$ decreases for all the values of $\delta$ for $\delta < 2$. Whereas in Fig. 2, $\rho_M$ is clipping 
through the progression of a model for $\beta$ and $\gamma$. Surprisingly we noticed that the energy density of THDE does not evolve for 
enough long values of time, but the energy density of matter disappears through the evolution. The compelling consequences of this research 
close that the lowering of the energy density of THDE for cosmic time crosses the volume extension of the Universe. We plotted the parameter 
of overall density $(\Omega)$ in Fig. 3. We noticed that in late time overall density approaches to 1. The research foretells that the anisotropy 
will damp out, also the Universe will be isotropic at large time. This result is considered the same as Samanta and Mishra \cite{ref99}, where 
they resemble that Universe is isotropy for an amply wide time. The density parameter shows a comparable form to this case for $\delta = 1.02$. 
Fig. 4 displays that the EoS parameter increases with time and converges to  $\omega_T \to -1$. EoS parameter of the research diversifies 
the quintessence field (-1$< \omega_T < -1/3$) through its evolution. The EoS parameter of this research resembles $\Lambda$CDM model 
($\omega_T = -1$) in eternity. The form has a more obvious accelerating impact that intimates the little charges of cosmic time. 
In Fig. 5, the deceleration parameter reduces from positive to negative zone plus eventually directs to $-1$. The model displays a change 
from decelerated phase to accelerated phase. Presently, for bigger values of $\Delta$, the model enters into an accelerated phase.\\

We have shown the conformity among the models of scalar field quintessence and the THDE. Fig. $6$ shows the plot of the scalar field ($\phi$) 
and potential ($V(\phi)$) at variance with $t$. The scalar field dissolves for liberal values of cosmic time and shows a decreasing function, 
whereas the potential converges to $2$ as t$\to$ $\infty$ and explains the decreasing function of time. \\

We displayed some of the modified distances as Distance modulus, Angular-diameter distance, and Luminosity distance in this research to know 
more about the Universe (Figs. $7$, $8$, and $9$). These results are found to be compatible with recent observations. Thus, our developed model 
and its solutions are physically acceptable.

\section*{Acknowledgments}
The authors thank the IUCAA, Pune, India for providing the facility during a visit where a part of this work was completed.

\end{document}